\documentclass[preprint,aps,showpacs]{revtex4}

\usepackage{graphicx}

\newcommand{\mgb}{MgB$_2$}
\newcommand{\hirr}{$H_{\rm irr}$}
\newcommand{\hc}{$H_{\rm c2}$}
\renewcommand{\deg}{$^{\circ}$}
\newcommand{\Tc}{$T_{\rm c}$}

\begin{document}

\title{
Enhancement of the irreversibility field by carbon substitution 
in single crystal MgB$_2$
}

\author{
E. Ohmichi$^1$, T. Masui$^{2,3}$, S. Lee$^{2}$,  S. Tajima$^2$, 
and T. Osada$^1$
}

\affiliation{$^1$Institute for Solid State Physics, University of Tokyo, Kashiwa 277-8581 Japan}

\affiliation{$^2$Superconductivity Research Center, ISTEC, 1-10-13 Shinonome, Koto-ku, Tokyo 135-0062 Japan}

\affiliation{$^3$Domestic Research Fellow, Japan Society for the Promotion of Science, Kawaguchi, Japan}

\begin{abstract}
We report the detailed study of the irreversibility field {\hirr} 
of single crystals of Mg(B$_{1-x}$C$_{x}$)$_2$ 
($x$=0, 2, 3.5, 5 and 7\%) in strong magnetic fields of up to 40 T. 
High sensitive torque measurements revealed that 
the {\hirr} is greatly enhanced by a factor of two 
($\mu_{0}${\hirr}(0)$\simeq$33 T in $x$=5\%) by carbon substitution 
compared to the pristine {\mgb} owing to a reduction 
in the inplane coherence length.  
We also observed the temperature-dependent {\hirr} 
anisotropy in all the carbon contents. 
This fact strongly suggests that the two-gap superconductivity of {\mgb} 
is less influenced by a small amount of carbon substitution. 
We discuss these results focusing on the impact of 
carrier doping and impurity scattering on two-gap superconductor. 

\end{abstract}

\maketitle

It is widely accepted that {\mgb} \cite{nagamatsu} is a two-gap 
superconductor based upon numerous experimental results 
\cite{szabo,iavarone,eskildsen,bouquet,angst,quilty,tsuda,souma}. 
In {\mgb}, two different order parameters reside on 
two types of distinct Fermi surfaces (FS) \cite{kortus} 
originating from two-dimensional (2D) $\sigma$ and 
three-dimensional (3D) $\pi$ bands, respectively. 
The $\sigma$  band, strongly coupled with the inplane $E_{2g}$ phonon mode, 
exhibits a larger superconducting gap than the $\pi$ band. 
The disparity between both bands suppresses impurity interband scattering 
that causes pair-breaking in a multigap superconductor, and thus 
two superconducting gaps with different energies are 
retained, even if the sample quality is not so high \cite{mazin}. 

Carbon substitution for boron in {\mgb} \cite{takenobu,ribeiro,lee}
is of great interest from the 
viewpoint of both carrier doping and impurity scattering 
in a multigap superconductor. 
Since carbon has one more electron than boron, it is expected that 
electrons are doped into the system by carbon substitution. 
This indicates that the hole $\sigma$ band dominating 
the superconductivity in {\mgb} is filled up, and thus 
the superconducting properties would be substantially modified \cite{singh}. 
On the other hand, carbon substitution brings about inplane scattering. 
As mentioned above, if the interband impurity scattering rate is 
substantially increased , 
the two-gap model is no longer valid \cite{golubov,ohashi}. 
Experiments with controlled impurity contents would be a critical test 
for the two-band model. 
Another interest in introduction of scattering centers is the enhancement of 
the upper critical field as usually seen in conventional superconductors. 
The improved superconducting properties are important 
from the viewpoint of practical application. 

In the present study, we performed 
high sensitive torque ($\tau$) measurements with high quality single crystals 
of carbon-substituted {\mgb} in fields of up to 40 T and at 
low temperatures down to 0.5 K. 
In order to investigate the detailed physical properties 
of anisotropic materials such as {\mgb}, single crystals are indispensable. 
In this paper, 
we report the detailed study of the irreversibility field {\hirr}, 
one of the most fundamental superconducting properties, of 
Mg(B$_{1-x}$C$_x$)$_2$ ($x$=0-7.5\%). 
We find that {\hirr} is enhanced by a factor of two in carbon-substituted 
sample, exceeding 30 T even at 4.2 K. 
We also observed 
a monotonous decrease in the {\hirr}-anisotropy 
$\gamma$ (={\hirr}$^{\|ab}$/{\hirr}$^{\|c}$) with carbon content as well as 
the temperature-dependent $\gamma$ in all the carbon contents. 
We discuss these results focusing on the impact of carrier doping and 
impurity scattering on the two-gap superconductor {\mgb}.


High quality single crystals of carbon-substituted {\mgb} 
were prepared by a high pressure technique \cite{lee}.
Carbon contents were determined by Auger electron spectroscopy. 
Samples were of plate-like shape with shiny surfaces (the $ab$ plane) 
with typical dimensions of $\sim$200$\times$200 $\mu$m$^2$. 
In this torque study, we have measured a set of Mg(B$_{1-x}$C$_x$)$_2$ 
($x$=0, 2, 3.5, 5, and 7.5\%) 
(hereafter denoted as C0\%, C2\%, C3.5\%, and so on). 
With increasing carbon content, 
the superconducting transition temperature {\Tc} monotonically decreased: 
{\Tc}=39 K in C0\% sample and {\Tc}=20 K in C7.5\% sample. 
Besides,  
the in-plane residual resistivity substantially increased more than 
by an order of magnitude in C7.5\% sample compared to C0\% sample.  
This is because boron atoms in the conducting layer 
are randomly replaced with carbon atoms. 

High field torque measurement was performed 
with use of a commercial piezoresistive microcantilever \cite{ohmichi}
(inset of Fig. 1). 
A sample is glued at the end of a cantilever beam 
with a small amount of epoxy. 
Dimensions of the microcantilever are 120$\times$50$\times$5 $\mu$m$^3$, 
thus enabling us to measure very small samples such as {\mgb}. 
A compensation lever is incorporated in the same platform for the purpose of 
eliminating temperature drift and magnetoresistance effect of piezoresistor. 
The eigenfrequency of 250-300 kHz is high enough to use 
in combination with a long pulse magnet with duration of 60 ms 
(maximum field of 42 T). 
The microcantilever was mounted on a rotating stage, and field orientation 
was changed with resolution of 1{\deg}. 
The polar angle $\theta$ is defined as the one tilted from the $c$ axis.


Figure 1 shows the typical behavior of torque divided by field ($\tau/H$) 
of C0\% sample at $\theta$=75{\deg} 
at different temperatures. 
Hysteretic behavior characteristic for type-II superconductor 
was observed below {\Tc}. 
We define the irreversibility field {\hirr} as 
the field where the hysteresis vanishes. 
The {\hirr} determined in this way coincides with  
the field $H_{\rho\rightarrow0}$ where the resistivity becomes zero. 
In this paper, we define {\hc} as the transition onset field and 
distinguish from {\hirr}. 
At elevated temperatures, the hysteresis becomes less pronounced, and thus 
the ambiguity in determining {\hirr} is inevitably large near {\Tc}. 
It is worth noting that the peak effect is clearly observed 
slightly below {\hirr} as reported \cite{angst2}. 
The peak amplitude becomes small with temperature, and it is almost 
invisible above 25 K. The detailed analysis will be published elsewhere. 
 
 In determining the {\hirr} in the highly symmetric directions 
such as $H\|c$ axis or $H\|ab$ plane, 
field orientation was set to be slightly off the direction 
(typically 1-2{\deg}). 
This is because when magnetic field is aligned to the principle axis,  
torque signals vanish in principle. 
The systematic error originating from this field misalignment 
is estimated to be less than 0.1\%, since the anisotropy $\gamma$ 
of carbon-substituted {\mgb} is modest (as discussed later, $\gamma\simeq$2-6) 
compared to high-{\Tc} cuprates giving $\gamma\sim$10-50.

Shown in Fig. 2 are the main results of this paper, 
the temperature dependence of {\hirr} in field orientations 
of $H\|ab$ and $H\|c$ at each carbon content. 
In C0\% sample, the extrapolated $\mu_{0}${\hirr}$^{\|ab}$(0) and 
$\mu_{0}${\hirr}$^{\|c}$(0) are 
16.4 and 2.9 T, respectively, both of which are in agreement 
with the previously reported values of single crystal 
\cite{zehetmayer,welp}. 
Surprisingly, C2\% and C5\% samples exhibit much higher {\hirr} in both 
$H\|ab$ and $H\|c$ directions, though {\Tc}'s are 
substantially decreased by carbon substitution: 
$\mu_{0}${\hirr}$^{\|ab}$(0)= 26.8 T and $\mu_{0}${\hirr}$^{\|c}$(0)= 5.5 T 
in C2\% sample ({\Tc}=35 K) 
and $\mu_{0}${\hirr}$^{\|ab}$(0)=33.4 T and $\mu_{0}${\hirr}$^{\|c}$(0)=8.1 T 
in C5\% sample ({\Tc}=27 K). 
With further increasing carbon content to 7.5\% ({\Tc}=20 K), 
$\mu_{0}${\hirr}$^{\|ab}$(0) decreased to $\sim$20 T  
while $\mu_{0}${\hirr}$^{\|c}$(0) still increased to $\sim$9 T \cite{C75}. 
This indicates that the zero-temperature anisotropy 
$\gamma$(0)={\hirr}$^{\|ab}$(0)$/${\hirr}$^{\|c}$(0) is reduced to about 2 
compared to $\gamma$(0)=5.5 in C0\% sample. 
Figure 3 summarizes the experimental results shown in Fig. 2. 
Both {\hirr}$^{\|ab}$(4.2K) and {\hirr}$^{\|c}$(4.2K) 
exhibit a maximum at around 3.5\%-carbon doping (Fig. 3(a)). 
So far, much effort has been devoted to increase the {\hirr} for 
practical application. 
The present data give a high {\hirr} value greater than 30 T 
even at 4.2 K in bulk single crystals, 
which exceeds the $\mu_{0}${\hc}(0)$\sim$29 T of Nb$_3$Sn \cite{orlando}. 
{\Tc} and $\gamma$ monotonously decrease with carbon 
content in a similar manner (Fig. 3(c)). 
In the following, we explain these systematic changes
from the viewpoint of carrier doping and impurity scattering caused by 
carbon substitution. 

First, we discuss the enhancement of {\hirr} by a factor of two 
by carbon substitution. 
To understand this, the change of coherence lengths becomes essential. 
Shown in Fig. 3(b) are the inplane and interplane coherence length 
$\xi_{ab}$(4.2K) and $\xi_{c}$(4.2K) derived from {\hirr} 
according to the relations 
$\mu_{0}${\hc}$^{\|c}$=$\Phi_{0}/2\pi\xi_{ab}^2$ and 
$\mu_{0}${\hc}$^{\|ab}$=$\Phi_{0}/2\pi\xi_{ab}\xi_{c}$ 
by assuming {\hirr}={\hc}.
Since {\hirr}$<${\hc} in a real material, particularly for $H\|c$ 
\cite{eltsev,masui}, 
the estimated $\xi_{ab}$ above is regarded as the uppermost value 
of the inplane coherence length. 
With increasing carbon content, the resistive transition width becomes 
narrower, and the broadening effect is small even in $H\|c$ 
\cite{masui2}.  
Therefore, the error due to the difference between {\hirr} and {\hc} 
is smaller in the higher carbon content region.
The radical increase in {\hirr} may be partly due to the reduction 
in the difference {\hc}$-${\hirr}

It is found in Fig. 3(b) that $\xi_{ab}$ monotonically decreases 
($\xi_{ab}$(4.2K)=10.5 nm in C0\% sample and  
$\xi_{ab}$(4.2K)=6 nm in C7.5\% sample), 
while $\xi_c$ is almost constant at lower carbon content, and is slightly 
increasing at higher carbon content. 
This drastic change in $\xi_{ab}$ indicates 
that carbon substitution has a strong influence on inplane properties 
as expected. 
A simple calculation gives that 
the inplane mean free path $\ell_{ab}$ for C0\% sample reaches several ten nm, 
while it is reduced down to several nm for C7.5\% sample owing to 
the increased impurity scattering rate \cite{masui2}. 
This fact strongly suggests that a crossover from a clean superconductor 
to a dirty one occurs with carbon content.  
Therefore, it is considered that 
{\hirr} is enhanced due to 
the reduced coherence length according to 
the well-known relation $\xi_{ab}^{-1}=\xi_{0ab}^{-1}+\ell_{ab}^{-1}$ where 
$\xi_{0ab}$ is the inplane coherence length without impurity scattering. 
Since {\hirr} also depends on {\Tc} 
(approximately {\hirr}$\propto${\Tc}), further carbon substitution,  
which decreases {\Tc}, leads to manifestation of a 
maximum as shown in Fig. 3(a). 

It is not straightforward to explain the reduction in $\gamma$(0), 
because both of impurity scattering and electron 
doping may affect $\gamma$(0). The increase of the 
interband impurity scattering rate $\Gamma_{\rm inter}$ 
wipes out two-gap superconductivity 
due to mixing of 2D $\sigma$ and 3D $\pi$ bands, so that 
we expect the reduced $\gamma$(0). 
However, the change of FS topography accompanied by doping is also 
directly reflected in $\gamma$(0) ($\it e.g.$ $\gamma$(0) is given by 
a square root of effective mass anisotropy in a one-gap superconductor).  
To answer the question what is the main mechanism 
for the reduction in $\gamma$(0), 
the temperature dependence of $\gamma$ in carbon-substituted samples 
is a critical test 
to evaluate the amplitude of interband impurity scattering rate 
$\Gamma_{\rm inter}$. 
In a pure {\mgb}, the contribution of the 3D $\pi$-band superconductivity 
is strongly suppressed in high magnetic fields, which result in a 
rapid increase in $\gamma$ with lowering temperature, followed by a 
saturation at low temperatures \cite{bouquet2}. In a naive picture,  
if $\Gamma_{\rm inter}\ll \Gamma_{\rm intra}$, where $\Gamma_{\rm intra}$ is 
the intraband impurity scattering rate, the system is still described by 
a two-gap model 
where $\gamma$ would be temperature dependent as in C0\% sample, 
while  if $\Gamma_{\rm inter}\approx \Gamma_{\rm intra}$, 
then the system would 
behave like a one-gap superconductor with temperature-independent $\gamma$. 

Figure 4 shows the temperature dependence of the anisotropy $\gamma$ 
in each carbon content. 
Since ambiguity of {\hirr} is not negligible in the high temperature region, 
the $\gamma$'s obtained from the resistivity measurements 
are plotted together. 
We confirmed that both of the torque and resistivity data give 
consistent $\gamma$ values in the intermediate temperature region. 
In C0\% sample, 
our data showed significant temperature-dependent $\gamma$ values 
ranging from 2 to 5.5, which is consistent with the previous reports 
\cite{angst,lyard}. 
With increasing carbon content, 
the value monotonously decreases, but still 
$\gamma$ is temperature-dependent in all the carbon contents 
($\gamma$(C2\%)=2-5, and $\gamma$(7.5\%)=1-2). 
This results strongly suggest that the two-gap superconductivity picture 
is still valid in carbon-substituted samples, and 
that the smaller $\pi$-band gap is robust against carbon substitution 
in contrast to an appreciable decrease in the larger $\sigma$-band gap. 
Thus, we conclude that the situation 
$\Gamma_{\rm inter}\ll \Gamma_{\rm intra}$ holds for small carbon contents. 
Therefore, the mechanism of the reduction in $\gamma$(0) 
is predominantly attributable to the effect of carrier doping. 
The evidences for electron doping are 
found in Hall coefficient \cite{masui2} and 
photoemission spectroscopy \cite{tsuda2} measurements indeed. 

As demonstrated in the specific heat \cite{bouquet2} and 
discussed theoretically in Ref. \cite{miranovic,dahm}, 
$\gamma$ in the low temperature region is dominated 
by the anisotropy of the $\sigma$ band. 
So, the reduced anisotropy $\gamma$(0) 
indicates more three-dimensional $\sigma$ FS.
According to the band calculation \cite{kortus}, the $\sigma$ band 
consists of a pair of quasi two-dimensional hole cylinders.  
When electrons are doped into a hole cylindrical FS,  
it gradually shrinks, then becomes an ellipsoidal FS, and is 
finally filled up completely. 
Therefore, we consider that the reduction in $\gamma$(0) primarily reflects 
the doping process into the $\sigma$ band. 
This view is supported by the fact that C12\% sample shows completely 
isotropic behavior \cite{lee}.  

Finally, we briefly mention the anomalous behavior of {\hirr}-$T$ curves, 
especially in $H\|ab$ (Fig. 2). 
{\hirr}(C0\%) and {\hirr}(C2\%) show conventional 
negative curvatures, but 
{\hirr}(C5\%) grows almost linearly, and  {\hirr}(C7.5\%)
even exhibits a positive curvature down to the lowest temperature. 
Since the transition broadening is small for $H\|ab$ and 
becomes small both for $H\|ab$ and $H\|c$ in the high 
carbon contents, similar behavior is also expected in {\hc}.  
Here we remind that a positive curvature of {\hirr} was observed at 
temperature very close to {\Tc} even in a pure {\mgb} 
\cite{sologubenko}. As is 
theoretically proposed \cite{gurevich}, 
it reflects the fact that 
{\hc} is dominated by the $\pi$-band gap near {\Tc}, while, 
at $T<${\Tc}, the $\sigma$ band determines {\hc}. It is likely that 
the carbon substitution introduces scattering also in the $\pi$ band and 
makes the $\pi$-band gap {\it more} robust against magnetic field. 
The change of {\hirr}-$T$ curves observed in Fig. 2 can be understood as 
the evidence that carbon substitution widens the temperature range 
where {\hirr} is affected by the $\pi$ band.

In conclusion, 
we have shown the enhancement of the irreversibility field {\hirr} at $T$=0 
by a factor of two by carbon substitution in single crystal of {\mgb}, 
although the transition temperature {\Tc} monotonically decreases 
with substitution. 
The enhancement is explained by the crossover from the clean 
to the dirty limit regime where the inplane mean free path $\ell_{ab}$ is 
less than the coherence length $\xi_{0ab}$ 
due to increased impurity scattering caused by carbon substitution. 
The reduced anisotropy in {\hirr} is ascribable to the 
reduced two-dimensionality of the $\sigma$ band 
as a result of electron doping. 
We also observed temperature-dependent anisotropy in all the carbon 
contents (C=0-7.5\%), which strongly suggests that 
the two-gap superconductivity model is still valid 
in these carbon-substituted {\mgb}. 

\begin{center}
{\small ACKNOWLEDGMENT}
\end{center}

We acknowledge F. Sakai for technical support and thank the 
Materials Design and Characterization Laboratory, Institute for Solid 
State Physics. 
This work was supported by New Energy and Industrial Technology 
Development Organization (NEDO) as Collaborative Research and Development 
of Fundamental Technologies for Superconductivity, and also 
by a Grant-in-Aid for Encouragement of Young Scientist 
from the Japan Society for the Promotion of Science.

\newpage

\begin{figure}
\includegraphics[width=12cm]{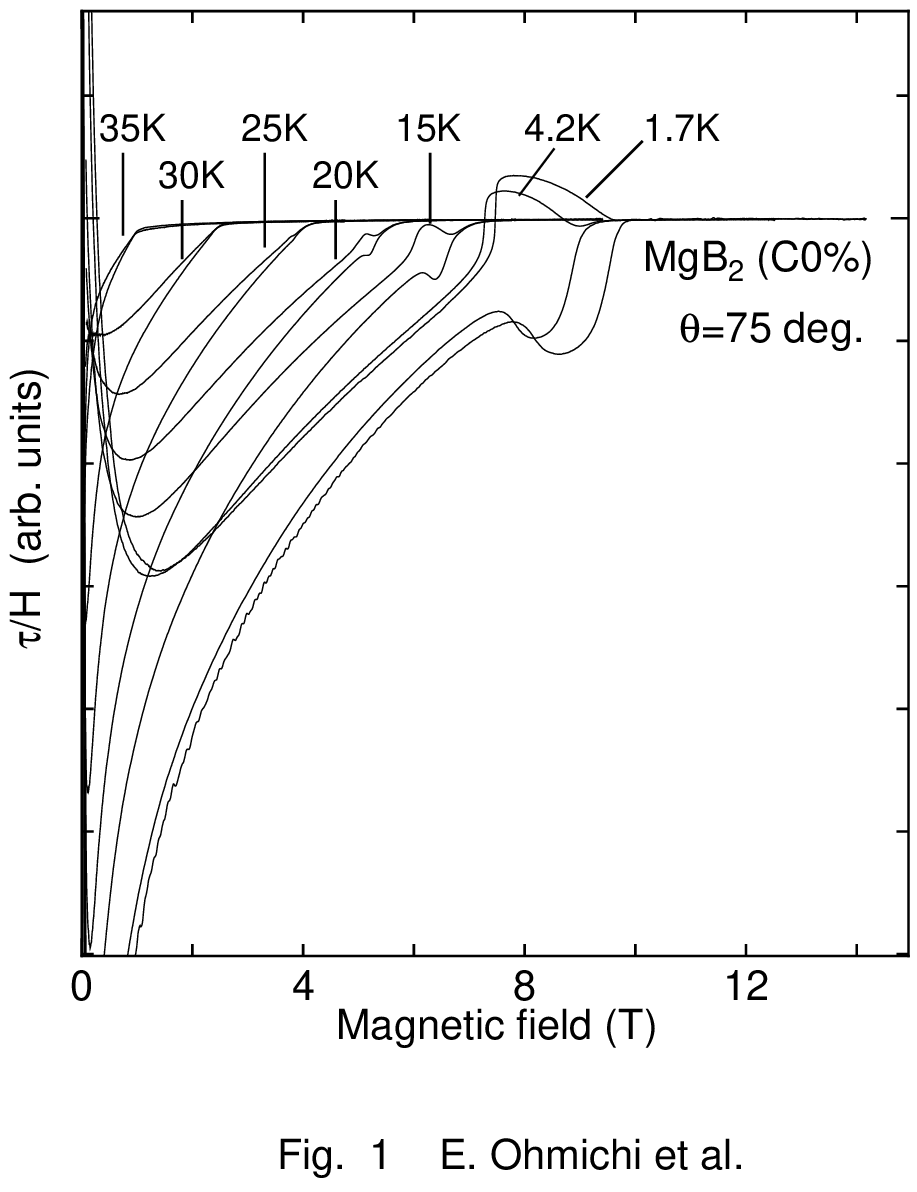}
\hspace{3.7cm}
\vspace{-8.7cm}

\includegraphics[width=70mm,height=53mm]{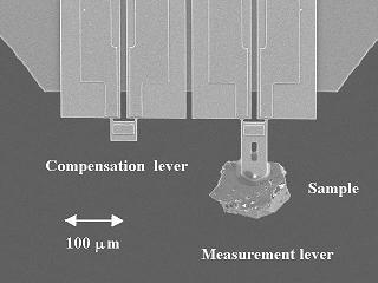}

\vspace{4cm}
\caption{
Typical behavior of torque divided by field ($\tau /H$) 
in C0\% sample at $\theta$=75{\deg} at 
different temperatures, where $\theta$ denotes the angle measured from 
the $c$ axis. Inset: scanning electron microscope image 
of a microcantilever. A {\mgb} single crystal is attached at the end of 
the cantilever beam.}   
\end{figure}

\begin{figure}
\includegraphics{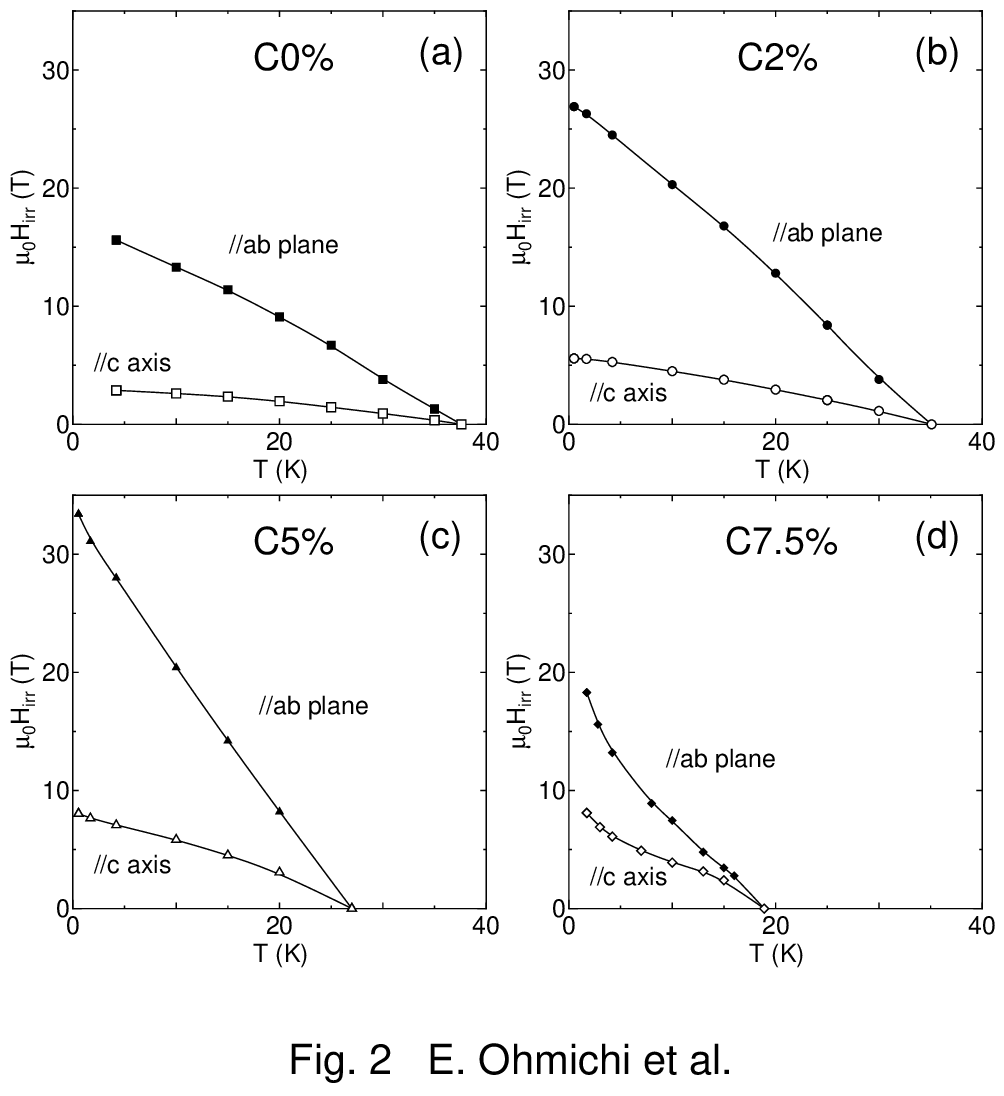}
\caption{
Temperature dependence of {\hirr}$^{\|ab}$ and {\hirr}$^{\|c}$ 
in (a) C0\%, (b) C2\%, (c) C5\%, and (d) C7.5\% samples. 
}
\end{figure}

\begin{figure}
\includegraphics{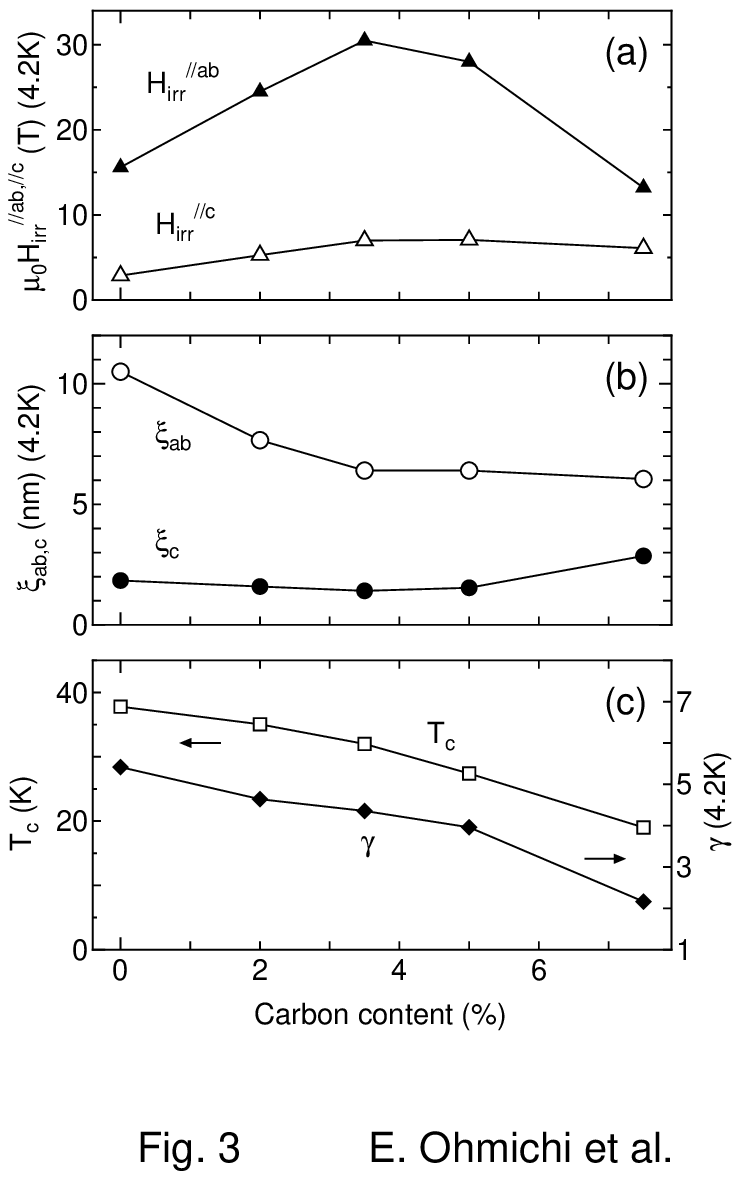}
\caption{
(a) The irreversibility field {\hirr}$^{\|ab}$, {\hirr}$^{\|c}$ at 4.2 K, 
(b) the coherence length $\xi_{ab}$, $\xi_{c}$ at 4.2 K, and 
(c) the transition temperature {\Tc} and the {\hirr} anisotropy $\gamma$
at 4.2 K as a function of carbon content. 
 }
\end{figure}

\begin{figure}
\includegraphics{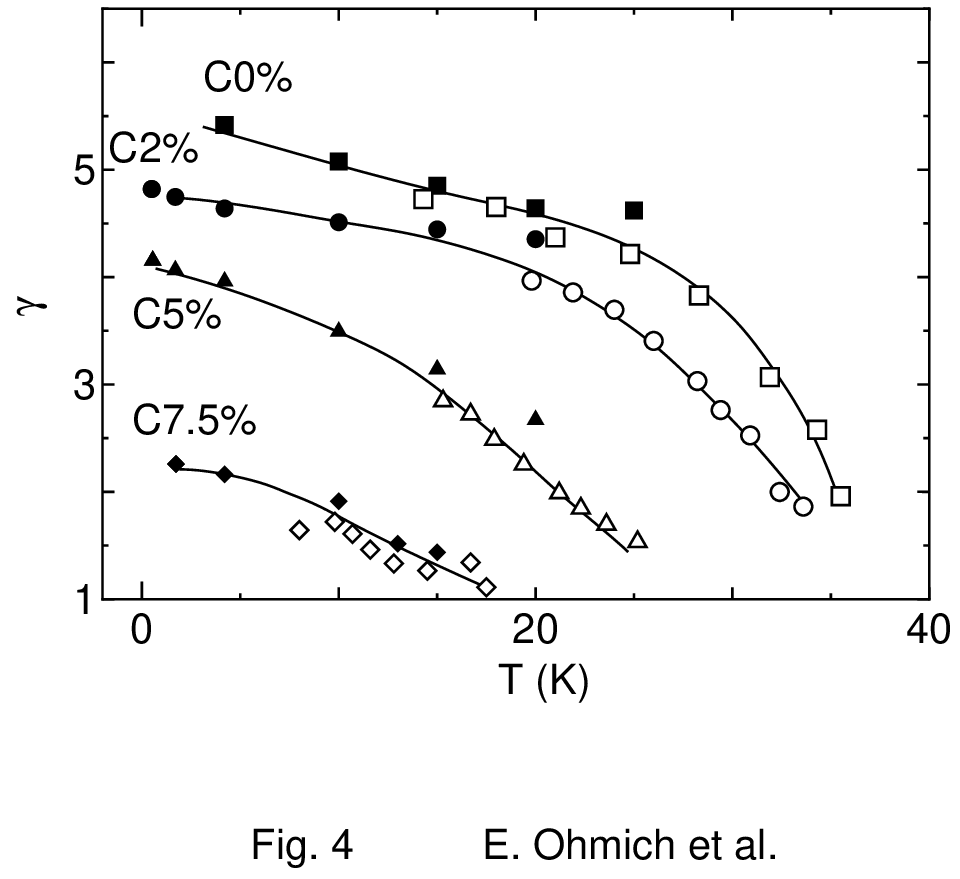}
\caption{
Temperature dependence of $\gamma$ in C0\%, C2\%, C5\%, and C7.5\% 
samples. The data shown by open and solid symbols are 
obtained by resistivity and torque measurement, respectively.  
The solid lines are guides for the eye. 
}
\end{figure}


\begin{references}

\bibitem{nagamatsu} 
J. Nagamatsu {\it et al.}, Nature {\bf 410}, 63 (2001). 

\bibitem{szabo}
P. Szab{\'o} {\it et al.}, Phys. Rev. Lett. {\bf 87}, 137005 (2001).  

\bibitem{iavarone}
M. Iavarone {\it et al.}, Phys. Rev. Lett. {\bf 89}, 187002 (2002). 

\bibitem{eskildsen}
M. R. Eskildsen {\it et al.}, Phys. Rev. Lett. {\bf 89}, 187003 (2002). 

\bibitem{bouquet}
F. Bouquet {\it et al.}, Phys. Rev. Lett. {\bf 87}, 047001 (2001).

\bibitem{angst}
M. Angst {\it et al.}, Phys. Rev. Lett. {\bf 88}, 167004 (2002).

\bibitem{quilty}
J. M. Quilty {\it et al.}, Phys. Rev. Lett. {\bf 90}, 207006 (2003).

\bibitem{tsuda}
S. Tsuda {\it et al.}, Phys. Rev. Lett. {\bf 91}, 127001 (2003).

\bibitem{souma}
S. Souma {\it et al.}, Nature {\bf 423}, 65 (2003). 

\bibitem{kortus}
J. Kortus {\it et al.}, Phys. Rev. Lett. {\bf 86}, 4656 (2001). 

\bibitem{mazin}
I. I. Mazin {\it et al.}, {Phys. Rev. Lett.}, {\bf 89}, 107002 (2002).

\bibitem{takenobu}
T. Takenobu {\it et al.}, Phys. Rev. B {\bf 64}, 134513 (2001).

\bibitem{ribeiro}
R. A. Ribeiro {\it et al.}, Physica C {\bf 384}, 227 (2003).

\bibitem{lee}
S. Lee {\it et al.}, Physica C {\bf 397}, 7 (2003).

\bibitem{singh}
P. P. Singh, Solid State Commun. {\bf 127}, 271 (2003).

\bibitem{golubov}
A. A. Golubov and I. I. Mazin, Phys. Rev. B {\bf 55}, 15146 (1997). 

\bibitem{ohashi}
Y. Ohashi, J. Phys. Soc. Jpn. {\bf 71}, 1978 (2002).

\bibitem{ohmichi}
E. Ohmichi and T. Osada, Rev. Sci. Instrum. {\bf 73}, 3022 (2002).


\bibitem{angst2}
M. Angst {\it et al.}, Phys. Rev. B {\bf 67}, 012502 (2003).

\bibitem{zehetmayer}
M. Zehetmayer {\it et al.}, Phys. Rev. B {\bf 66}, 052505 (2002). 

\bibitem{welp}
U. Welp {\it et al.}, Physica C {\bf 385}, 154 (2003). 

\bibitem{C75}
The extrapolation in C7.5\% sample contains large ambiguity  
due to complicated temperature dependence. 

\bibitem{orlando}
T. P. Orlando {\it et al.}, Phys. Rev. B {\bf 19}, 4545 (1979).@

\bibitem{eltsev}
Yu. Eltsev {\it et al.}, Phys. Rev. B {\bf 65}, 140501 (2002).

\bibitem{masui}
T. Masui {\it et al.}, Physica C {\bf 383}, 299 (2003).

\bibitem{masui2}
T. Masui {\it et al.}, preprint. 

\bibitem{bouquet2}
F. Bouquet {\it et al.}, Phys. Rev. Lett. {\bf 89}, 257001 (2002).

\bibitem{lyard}
L. Lyard {\it et al.}, Phys. Rev. B {\bf 66}, 180502 (2002).


\bibitem{tsuda2}
S. Tsuda {\it et al.}, private communication.


\bibitem{miranovic}
P. Miranovi{\'c} {\it et al.}, J. Phys. Soc. Jpn. {\bf 72}, 221 (2003).

\bibitem{dahm}
T. Dahm and N. Schopohl, Phys. Rev. Lett. {\bf 91}, 017001 (2003).

\bibitem{sologubenko}
A. V. Sologubenko {\it et al.}, Phys. Rev. B {\bf 65}, 180505 (2002).

\bibitem{gurevich}
A. Gurevich, Phys. Rev. B {\bf 67}, 184515 (2003).

\end{references}
\end{document}